# Seismological study of meta-instability of the Yangbi $M_S$ 6.4 earthquake in 2021

Shouyong Li✉

*Handan Central Seismic Station, Handan Hebei* 056001, *China.*

**Key points:**

- This paper describes the direction, magnitude, and intensity of the stress field from the perspective of vectors, using the gradient and the divergence of the gradient of stress factors, thereby reflecting the stress field.
- The consistency between the gradient and the gradient divergence of the apparent wave velocity ratio and apparent Poisson's ratio verify the experimental relationship between wave velocity and stress factors.
- The consistency between the distribution of the apparent wave velocity ratio and apparent Poisson's ratio values and their gradient divergence reflects the mutual interpretation and consistency of different physical fields in the meta-instability stage and also highlights the uniqueness of its precursors.
- The nonlinear relationship between apparent wave velocity ratio and apparent Poisson's ratio is demonstrated by the consistency of their gradient divergence, which also confirms the objectivity of their relationship.
- The analysis of the meta-instability of the Yangbi $M_S$6.4 earthquake indicates that the earthquake had precursors, and the results obtained from different observation methods are consistent.[1]
- This paper describes the convergence and divergence of energy at a certain points in the stress field.

**Abstract**: Meta-instability is an irreversible precursor of earthquakes. To identify the meta-instability precursor of the Yangbi $M_S$6.4 earthquake (99.87°E、25.67°N)that occurred on May 21, 2021, we selected seismic data from the pre-earthquake period between 1 and 21 May. We then calculated the apparent wave velocity ratio and the apparent Poisson's ratio within the region of 98.5°E-101°E，24.6°N-27.1°N and interpolated these values. Our findings revealed that the trends of the fitted straight lines at the maximum and minimum points of the gradient divergences of the apparent wave velocity ratio and apparent Poisson's ratio fields are consistent with the source mechanism solution for Sections 1 and 2, respectively. Similarly, the trend of the fitted straight lines at the minimum and maximum points of their values is also consistent with the source mechanism solution for Sections 1 and 2. Positive gradient divergence values indicate energy released, whereas negative values suggest energy absorption. The observed stress state matches the experimentally demonstrated meta-instable state. We propose that this method can be a reference for identifying the meta-instability of strike-slip strong earthquakes with a significant number of foreshocks. For seismically active regions, increasing the number of stations with rich data acquisition will facilitate more convenient stress field analysis.

✉ **Corresponding author.** Li SY, email: lishy@sina.com.

# 1. Introduction

Earthquakes are the result of the movement of stress fields, and it is of great significance to closely monitor the stability of crustal media under the action of stress fields is of great significance. During the movement of stress fields, the crust exhibits four states: steady state, meta-stable state, meta-instable state, and unstable state. Jin Ma's team conducted an in-depth analysis of these states through experiments and pointed out that meta-instability is an irreversible precursor to earthquake rupture. This stage has rich precursor activity characteristics and has been provided with specific judgment criteria, this is of great significance for capturing earthquake precursors and studying earthquake prediction. To apply the theory of meta-instability to field observations, researchers have explored its characteristics from various perspectives, including water level (Zhang SL et al., 2021), deformation (Zhang SL et al., 2016, 2021), gravity (Guo SS et al., 2021), and temperature (Ren YQ, 2015; Chen SY et al., 2021), which provided corresponding descriptions of its features. Although meta-instability is a critical state before fault rupture under the action of the stress field, describing it from the perspective of the stress field may have a direct and obvious effect. However, these methods face the challenge of an invisible and untouchable stress field. Nevertheless, researchers have proposed a novel method: using other physical quantities to indirectly reflect the stress field. XH Liu and DL Lai (1986) discovered that abnormal wave velocity is primarily influenced by stress factors and fault structures through experiments, which provides a potential method for stress field research. Assuming that the medium remains steady, the state of wave velocity can be regarded as a direct indicator of stress field changes. Moreover, the relative stability of Earth's tectonic pattern provides feasibility (Ma J, 2009) for using seismic wave velocity to reflect the stress field. Although the discontinuity of observation points and the irregular distribution of epicenters pose challenges, interpolation techniques can effectively approximate the properties of the medium at the interpolation point, and analyze stress changes from a field perspective. Therefore, the relationship between seismic wave velocity and stress field, as well as the comprehensive application of interpolation methods, is an effective way to observe stress field changes and study meta-instable states.

On May 21, 2021, at 21:48:34, an $M_S$6.4 earthquake occurred in Yangbi, Yunnan Province, with the epicenter located at 99.87°E and 25.67° N. The numerous fore-shocks and main shock faults with shallow strike-slip rupture faults (Yang JY et al., 2021; Long F et al., 2021; Lei XL et al., 2021), are providing favorable conditions for capturing meta-instability information. As a retrospective study, seismic data were used to calculate the values of the apparent wave velocity ratio and apparent Poisson's ratio and interpolate them related to seismic wave velocity, to achieve grid continuity in the region. The gradient and gradient divergence were used to reflect the changes in the stress field. Combined with the distribution characteristics of their contour lines, the changes in the stress field before the earthquake were revealed. Their morphologies and intrinsic connections are aimed at capturing meta-instable stages and discovering precursor information in field observations, providing a reference for earthquake prediction.

# 2. Geologic background

The $M_S$6.4 earthquakes in Yangbi, Yunnan, occurred on the Yunnan-Burma Block, which is the main active framework formed by the compression of the southeastern margin of the Tibet Plateau. This compression results from the northward subduction of the Indian Plate beneath the Eurasian Plate (Deng QD et al., 2002, 2009; Zhang PZ et al., 2003, 2005; Zhang GM et al., 2005). The western boundary of this block is the arc-shaped subduction zone of Myanmar, while its eastern boundary is the Jinshajiang-Honghe Fault Zone (Zhang PZ et al., 2003). Since the Cenozoic era, particularly during the late Quaternary period, tectonic activity has been very intense. Figure 1(a) shows its geographical distribution, with the study area indicated by green squares and multiple fault zones within it. The distribution of these faults may reflect the historical activity patterns of stress fields and earthquake distribution. Figure 1(b) shows the distribution of major fault zones within the study area, with the fault zone data sourced from literature (Deng QD et al., 2007).

Weixi-Qiaohou Fault Zone: This fault zone extends from Tongxun to Qiaohou in a northwesterly trend of 20° to 30°. Its southern section dips northeast at an angle of about 50° and is approximately 280 km long (Shao ZG, 2003). Since the early Cenozoic era, the fault has exhibited right-lateral strike-slip characteristics, transitioning to left-lateral strike-slip features in the Upper Miocene (Chang ZF et al., 2016, 2021, 2022);

Chenghai Fault Zone: This fault zone formed during the Cambrian period, it is approximately 200 km long. Multiple fault basins have developed along the fault zone, and their nature is sinistral strike-slip and normal fault (Geological Research Institute of China Earthquake Administration, et al., 1990; Guo SM et al., 1988);

Honghe Fault Zone: This fault zone, originating in Eryuan, Yunnan in the north and extending southeastward into Vietnam near the river mouth in the southeast, is about 1000 km, with approximately 600 km within Yunnan. The northern end is a fault group with alternating strike-slip and dip-slip movements (Institute of Geology, China Earthquake Administration, et al., 1990; Tian P et al., 2023). Since the early Quaternary period, it has experienced large-scale right-lateral strike-slip movements, with the canyon section primarily characterized by right-lateral movements (Guo SM et al., 1988, 1996);

Lijiang-Xiaojinhe Fault Zone: This fault zone, located on the northwestern plateau of Yunnan, with a north-northeast trend of 40°, is approximately 360 km long. It is a high-angle reverse left-lateral strike-slip active fault (Xiang HF et al., 2002; Ding R et al., 2018; Gong Z et al., 2023);

Longpan-Qiaohou Fault Zone: This fault zone, starting from Longpan in the north and ending at Qiaohou in the south, is approximately 120 km long. Initially a thrust fault, it transformed into a left-lateral strike-slip movement after the Tertiary period. (Institute of Geology, China Earthquake Administration, et al., 1990; Chang ZF et al., 2023);

Heqing-Eryuan Fault Zone: This fault zone, originated from the Xiaojinhe Fault in the north and formed an arc-shaped structure near Eryuan, this features northeast-trending faults extending southeastward. Its two significant characteristics are left-lateral strike-slip and lateral extension (Geological Research Institute of China Earthquake Administration et al., 1990; Fang YG et al., 2019);

Lancangjiang Fault Zone: This fault zone, stretching along the west side of the Lancang river, starts from Chongshan in the north and extends southward through Damenglong into Myanmar. Since the Quaternary period, it has primarily exhibited right-lateral strike-slip characteristics (Li GX, 1994);

Nujiang Fault Zone: This fault zone, approximately 500 km long in western Yunnan, curves from north to south. As the fault curves, the dip angle of the section decreases, and the lateral angle of the stretching line increases. The northeast-trending section exhibits both strike-slip and push-over properties (Li JC, 1998).

## 3 Method and Rationale

### 3.1. Method

Meta-instability stage is the period from the accumulation of strain energy under the action of the stress field to the extreme point of rock strength to the rupture of faults, and it is also the most active stage of earthquake precursors. Therefore, from the perspective of stress field analysis, it should be the most direct and obvious method to identify meta-instability and discover precursors. In a stress field, the gradient of stress factors indicates the direction of the fastest stress change, which can be interpreted as the direction and magnitude of stress, while the divergence of the gradient reflects the density change rate of the 'stress line', thereby indicating changes in the strength of the stress field. Due to the invisibility of stress fields. Given the invisibility of stress fields, visualizing these two factors is crucial for studying stress fields. According to the experiment on the relationship between seismic wave velocity and stress factors (Liu XH and Lai DL, 1986), when the influence of small earthquakes on the stress field is not considered, the stress field is stable, that is, the occurrence of small earthquakes does not affect the trend of stress field convergence and divergence at that location. Therefore, the changes in the stress field correspond one-to-one with the changes in seismic wave velocity. This means that on a 'stress field line', there must be a unique 'P-wave velocity line' and a unique 'S-wave velocity line', as well as a unique 'wave velocity ratio line' and a unique 'Poisson's ratio line', whose spatial distribution is the same as that of the stress field. Similarly, their gradients can reflect the direction and magnitude of stress, and the divergence of gradients can reflect changes in the strength of the stress field. Considering the complexity of the medium and the stress field, this paper only addresses the strength of the stress field and the direction of the stress, rather than the magnitude of stress. Due to the inaccessible nature of the Earth's crust, stations cannot directly measure seismic wave velocities. Therefore, stress fields can be reflected by measuring apparent P-wave velocities, apparent S-wave velocities, apparent wave velocity ratios, and apparent Poisson's ratios. To effectively represent the stress field, the observation values of the seismic source at the vertical position on the surface are taken as the research data. To make the observed quantity have field characteristics, interpolation processing is performed to achieve grid continuity of numerical values. When the stress field is orthogonal to the surface, the gradient and gradient divergence of these quantities can effectively reflect the changes in the direction and intensity of the stress field within the depth range of the earthquakes. Therefore, analyzing their characteristics is a feasible method for identifying meta-instability and discovering precursor information.

## 3.2. Calculation

The apparent P-wave velocity and S-wave velocity are important physical quantities for studying earthquakes and crustal media. However, their calculation involves the depth of the earthquake source, which is subject to large errors. To eliminate the influence of this error, the apparent wave velocity ratio and apparent Poisson's ratio are used for research. As the calculation of apparent velocity ratio and apparent Poisson's ratio involves the values of apparent P-wave velocity and apparent S-wave velocity, the calculation formulas for apparent P-wave velocity, apparent S-wave velocity, apparent wave velocity ratio, and apparent Poisson's ratio (Christensen. 1996) are also provided here:

$$v_{apparent_p} = \frac{\iota}{t_{pd} - t_o}. \tag{1}$$

$$v_{apparent_s} = \frac{\iota}{t_{sd} - t_o}. \tag{2}$$

$$r = \frac{v_{apparent_p}}{v_{apparent_s}} = \frac{\frac{\iota}{t_{pd} - t_o}}{\frac{\iota}{t_{sd} - t_o}} = \frac{t_{sd} - t_o}{t_{pd} - t_o}. \tag{3}$$

$$\gamma = 0.5(1 - \frac{1}{r^2 - 1}) = 0.5(\frac{v^2_{apparent_p} - 2v^2_{apparent_s}}{v^2_{apparent_p} - v^2_{apparent_s}}). \tag{4}$$

Where: $\iota$ represents the focal distance, $t_{pd}$ represents the arrival time of the P-wave, $t_o$ represents the time of the earthquake, $t_{sd}$ represents the arrival time of the S-wave, $r$ represents the apparent wave velocity ratio, and $\gamma$ represents the apparent Poisson's ratio.

For regional grid division, the epicenter position is used to determine the grid, and the average or median of the apparent wave velocity ratio and apparent Poisson's ratio within the same grid is taken as the equivalent value of the grid. Then, interpolation is performed to form continuous grid values within the region, endowing them with field characteristics.

## 3.3. Stress direction and stress field strength

The gradient represents the direction of the fastest change in the apparent wave velocity ratio or apparent Poisson's ratio within the field, forming a vector field. The divergence of gradients represents the rate of density change of gradients per unit area in a field. The calculation of the gradient is shown in equation (5), where **i** and **j** represent unit losses. If the vector field of the gradient is $A = p(x,y)\mathbf{i}+q(x,y)\mathbf{j}$, its divergence calculation is shown in equation (6) (Xie SY, 2019).

$$\mathbf{grad}f = \frac{\partial f}{\partial x}\mathbf{i} + \frac{\partial f}{\partial y}\mathbf{j}. \qquad (5)$$

$$\mathrm{div}A = \frac{\partial p}{\partial x} + \frac{\partial q}{\partial y}. \qquad (6)$$

A high field density indicates a high intensity, a decrease in field density indicates flux outflow and an increase in field density indicates flux inflow. Flux can be defined as the energy in the gradient direction per unit area over a certain period. The divergence of a gradient is a measure of the change in the strength of the gradient field. A positive divergence value indicates a high field strength and the outward emission of energy; A negative divergence value indicates a low field strength, leading to energy accumulation from external sources. The direction of the gradient is the fastest direction from the area with lower density to the area with higher density, indicating the direction of energy flow, which is consistent with the direction of stress. When the stress field is orthogonal to the surface, the gradient and its divergence can effectively reflect the changes in stress direction and intensity of the stress field.

## 4. Data and results

### 4.1. Selection and processing of data

Because the period from May 1st to 21st, 2021, before the earthquake was characterized by a relatively independent phase of intense small earthquake activity, the data during this period was taken as the research focus. This study utilized the observation reports from China Earthquake Networks Center as the fundamental dataset. Seismic data within the geographical range of 98.5°E-101.0°E, and 24.6°N-27.1°N were extracted. The selection criteria included earthquakes with epicentral distances between 20 to 100 km, positioning residuals less than 0.5, and the presence of Pg and Sg waves recorded in all four quadrants around the epicenter. A total of 100 earthquakes meeting these criteria were selected, with an average residual value of 0.3029. The epicentral distribution and residual histograms are depicted in Figure 1 (b) and (c), respectively. For these selected events, the apparent wave velocity ratio and apparent Poisson's ratio were calculated. To mitigate noise and stabilize the interpolated data, the study area was divided into a 10×10 grid system. Within each grid, the median values of the apparent wave velocity ratio were computed and used as the representative values for preprocessing. Considering the following factors (i) the smoothness and stability advantages of spline interpolation; (ii) the continuous and differentiable characteristics of crustal media within a certain scale; and (iii) the uneven distribution characteristics of the epicenter location, the adjustable tensor continuous curvature spline interpolation method (Smith and Wessel, 1990) was employed. The data were finely processed with a 1m × 1m grid resolution to ensure the continuity of the grid values across the entire region, thereby endowing the dataset with field-like characteristics.

The standard deviation of the apparent wave velocity ratio on a 10×10 grid and the residual after grid interpolation are shown in Figure 2 (a) and (b), respectively. Similarly, the standard deviation of the apparent Poisson's ratio on a 10×10 grid and the residual after grid interpolation are shown in Figure 2 (c) and (d), respectively. Because there are no records of deep faults in this area, the data quality and processing methods are deemed reliable.

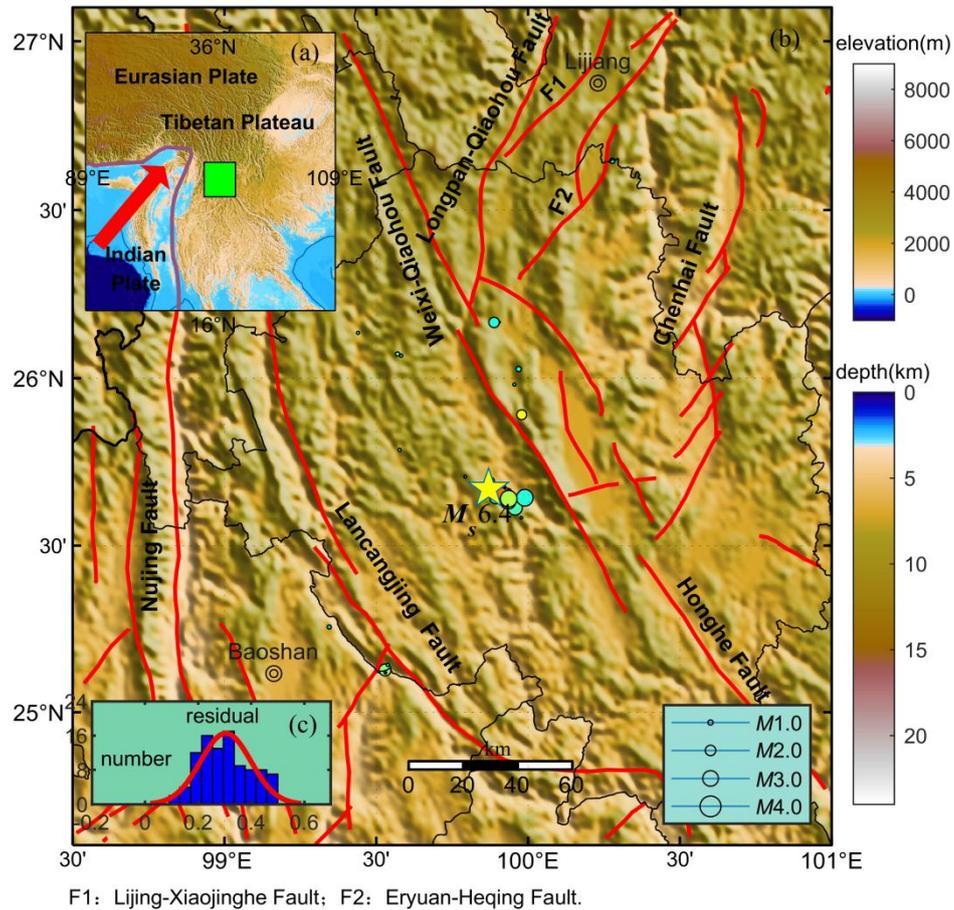

Fig.1 (a)Regional Location map. The plate boundary data is from literature(Bird,2003). The green square indicates the studied area; (b) Epicenter distribution map, the yellow star indicates the epicenter position of the $M_S$6.4 earthquakes, same for following the figures ; (c) Histogram of residuals.

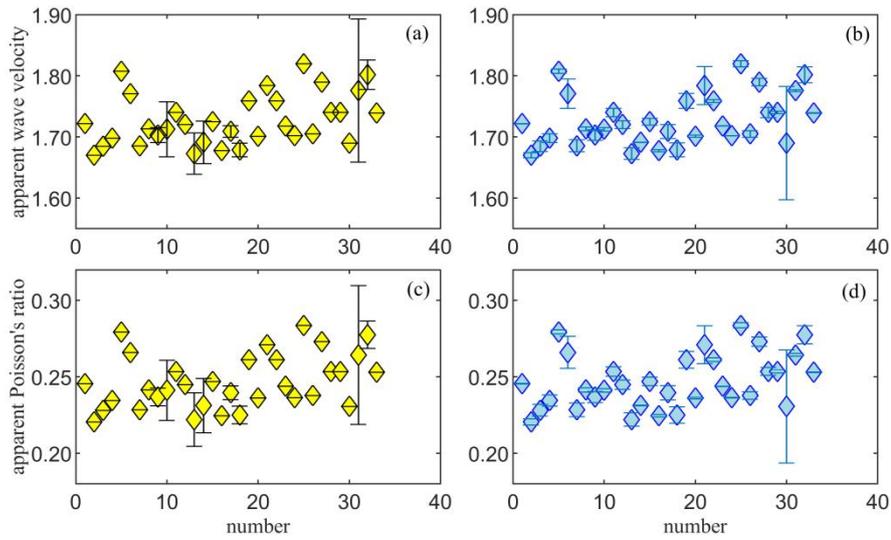

Fig.2 (a) Standard deviation of the median of the apparent wave velocity ratio in ten equal parts of the grid; (b)Residuals after interpolation of the apparent wave velocity ratio; (c) Standard deviation of the median of the apparent Poisson's ratio in ten equal parts of the grid; (d) Residuals after interpolation of the apparent Poisson's ratio.

### 4.2. Apparent wave velocity ratio

The contour lines of the apparent wave velocity ratio are shown in Figure 3 (a). The minimum points identified are A (99.56°E, 26.05°N) and B (99.84°E, 25.70°N), with a distance of 48.6 km between them. The trend of the fitted straight line a through these points is 142.8°. The maximum points are C (100.27°E, 26.66°N), D (99.87°E, 26.17°N), E (99.55°E, 25.76°N), and F (99.20°E, 25.42°N), with the trend of the fitted straight line b being 40.5°. The gradient and its divergence of the apparent wave ratio are shown in Figure 3(b). The gradient vectors indicate the direction of the stress field, while the divergence of the gradient reflects the degree of change in the stress field. The maximum points of divergence are A′ (99.60°E, 26.10°N) and B′ (99.88°E, 25.74°N), with a distance of 49.1 km between them. The trend of the fitted line a′ through these points is 141.8°. The minimum points of divergence are C′ (100.30°E, 26.59°N), D′ (99.98°E, 26.11°N), E′ (99.58°E, 25.79°N), and F′ (99.30° E, 25.30° N), with the trend of the fitted line b′ being 39.1°.

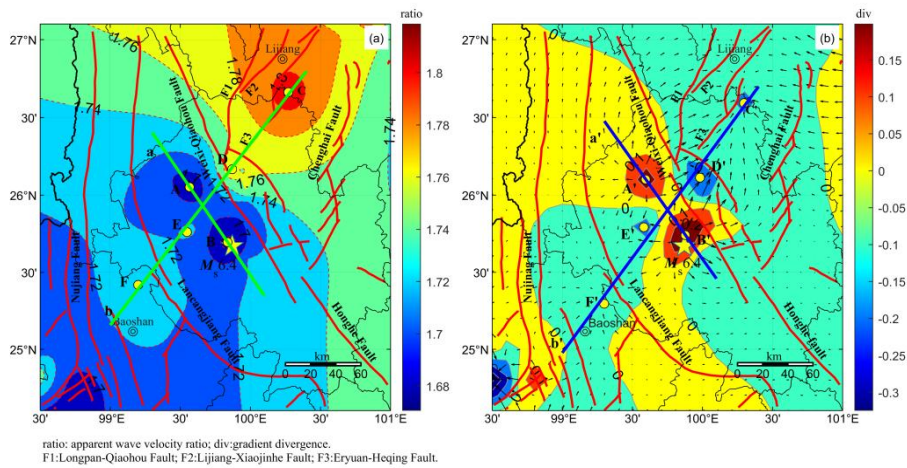

Fig. 3 (a) Distribution map of the Isolines of apparent wave velocity ratio. Points A and B are local minima, while points C, D, E, and F are local maxima. Line a is the fitting line through the local minima, and line b is the fitting line through the local maxima;(b) Distribution map of the gradient and gradient divergence of the apparent wave velocity ratio field. Points A´, and B´ are local maxima, while points C´, D´, E´,and F´ are local minima. Line a´ is the fitting line through the local maxima and line b´ is the fitted line through the local minima.

## 4.3. Apparent Poisson's ratio

The contour lines of the apparent Poisson's ratio are shown in Figure 4 (a). The minimum points identified are A (99.58°E, 26.05°N) and B (99.84°E, 25.70°N), with a distance of 46.9 km between them. The trend of the fitted straight line a through these points is 143.9°. The maximum points are C (100.28°E, 26.65°N), D (99.88°E, 26.16°N), and E (99.32°E, 25.25°N), with the trend of the fitted straight line b being 34.2°. The gradient and its divergence of the apparent Poisson's ratio are shown in Figure 4 (b). The gradient vectors indicate the direction of the stress field, while the divergence of the gradient reflects the degree of change in the stress field. The maximum points of divergence are A′ (99.60°E, 26.10°N) and B′ (99.87°E, 25.73°N), with a distance of 49.9 km between them. The trend of the fitted line a′ through these points is 144.0°. The minimum points of divergence are C′ (100.29°E, 26.60°N), D′ (99.98°E, 26.12°N), E′ (99.59°E, 25.79°N), and F′ (99.27°E, 25.48°N), with the trend of the fitted line b′ being 49.9°.

The above shows the fitting lines of the minimum and maximum points of the apparent wave velocity ratio and apparent Poisson's ratio, as well as the corresponding fitting lines of the maximum and minimum points of their respective gradient divergence, exhibit specific spatial patterns. Specifically, the minimum points of the apparent wave velocity ratio and apparent Poisson's ratio correspond to the fitted lines of the minimum points, the maximum points correspond to the fitted lines of the maximum points, while the fitted lines of the minimum points correspond to the fitted lines of the maximum points of the gradient divergence, and the fitted lines of the maximum points correspond to the fitted lines of the minimum points of the gradient divergence.

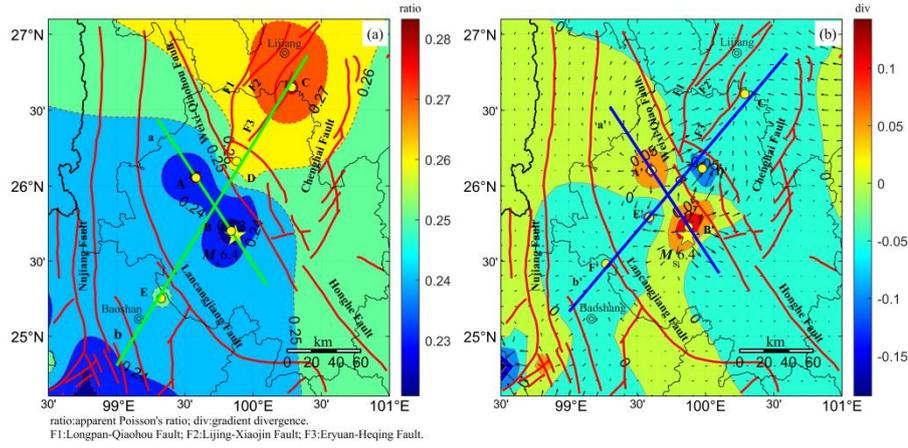

Fig. 4 (a) Distribution map of the Isolines of apparent Poisson's ratio. Points A and B are local minima, while points C, D, and E are local maxima. Line a is the fitting line through the local minima, and line b is the fitting line through the local maxima; (b) Distribution map of the gradient and gradient divergence of the apparent Poisson's field. Points A´ and B´ are local maxima, while points C´, D´ and E´are local minima. Line a´ is the fitting line through the local maxima and line b´ is the fitting line through the local minima.

## 5. Discussion

### 5.1. Analysis of apparent wave velocity ratio field

Figure 3 (a) shows the contour distribution of the apparent wave velocity ratio, with minimum value points A and B, and maximum value points C, D, E, and F. Figure 3 (b) shows the variation of the gradient and gradient divergence of the apparent wave velocity ratio field, the maximum value points A′ and B′ represent energy release, while the minimum value points C′, D′, E′, and F′ represent energy absorption. The characteristics exhibited by points A′, B′, C′, D′, E′, and F′ are consistent with the energy activity pattern in the meta-instability stage and reflect the degree of synergism on the fault (Ma J et al., 2012; Ma J and Guo YS, 2014; Ma J, 2016). These points show the activity traces of the fault. According to literature (Chen SY et al., 2009; Deng MD et al., 1997), it is known that the temperature at which rock media releases energy decreases while the temperature at which it absorbs energy increases. Temperature changes have a highly sensitive effect on the wave velocity ratio (Feng DY, 1981). As the temperature decreases, the wave velocity ratio rapidly decreases, while as the temperature increases, the wave velocity ratio rapidly increases. Experimental results show that (i) the wave velocity ratio decreases significantly with increasing pressure; (ii) the wave velocity slowly increases with increasing pressure; (iii) the velocity ratio slowly decreases with increasing pressure; (iv) there are four relationships of slow fluctuations with increasing pressure (Geng NG et al., 1992). Therefore, it is speculated that the main reason for the change in wave velocity ratio in the seismic source environment may be the temperature change of the medium caused by the release or absorption of energy during the stress field change process. This explains why the geographical locations of points A, B, C, D, E, and F correspond one-to-one with points A´, B´, C´, D´, E´, and F´. As the absorbed energy temperature increases, the apparent wave velocity ratio significantly increases, while the released energy temperature decreases and the apparent wave velocity ratio significantly decreases. The distance between the minimum points A and B shown in Figure 3 (a) is 48.6 km, and the trend of the fitted line is 142.8°. The distance between the maximum points A´ and B´ shown in Figure 3 (b) is 49.1 km, and their fitted line trend is 141.8°; The trend of the fitted line for the maximum points C, D, E, and F in Figure 3 (a) is 40.5°, while the trend of the fitted line for the minimum points C′, D′, E′, and F′ in Figure 3 (b) is 39.1°. Although Figure 3 (a) and Figure 3 (b) represent two different physical fields, they are essentially the same and interpretable. They are both manifestations of energy release and absorption on the locked fault during the meta-instability stage, except that the feature shown in Figure 3 (a) is indirectly caused.

## 5.2. Analysis of apparent Poisson's ratio field

Figure 4 (a) shows the contour distribution of apparent Poisson's ratio, with minimum value points A and B, and maximum value points C, D, and E. Figure 4 (b) shows the variation of the gradient and gradient divergence of the apparent Poisson's ratio field, the maximum value points A′ and B′ represent energy release, while the minimum value points C′, D′, E′, and F′ represent energy absorption. The characteristics reflected by A´, B´, C´, D´, E´, and F´ are consistent with the activity traces of the fault, indicating a certain degree of synergy. The energy activity pattern is also consistent with the characteristics of the meta-instability stage (Ma J et al., 2012; Ma J and Guo YS, 2014; Ma J, 2016). According to literature (Ji SC et al., 2009; Chen AG et al., 2019), changes in Poisson's ratio, similar to wave velocity ratio, are highly sensitive physical quantities to temperature changes. As the temperature decreases, Poisson's ratio rapidly decreases, and as the temperature increases, Poisson's ratio rapidly increases. However, once the confining pressure reaches a certain level, there will be no significant changes due to further changes in confining pressure. Like the apparent wave velocity ratio, it is speculated that the main reason for the change in apparent Poisson's ratio may be due to the release or absorption of energy causing temperature changes in the medium. This explains why the geographical locations of points A, B, C, D, and E correspond with points A´, B´, C´, D´, E´, and F´. As the absorbed energy temperature increases, the apparent Poisson's ratio significantly increases, while the released energy temperature decreases and the apparent Poisson's ratio significantly decreases. In Figure 4 (a), the distance between the minimum points A and B is 46.9 km, and the trend of the fitted line is 143.9°. In Figure 4 (b), the distance between the maximum points A' and B' is 49.9 km, and the trend of the fitted line is 144.0°. In Figure 4(a), the trend of the fitted line for the maximum points C, D, and E in Figure4(a) is 34.2°, and in Figure 4(b), the trend of the fitted line for the minimum points C´, D´, E', and F´ is 49.9°. Their geographical locations are slightly different, suggesting that this may be due to the concentration of energy release and the wide range of energy absorption, as well as the mismatch between thermal conductivity and energy release rate under complex medium conditions. Although Figure 4 (a) and Figure 4 (b) represent two different physical fields, they are essentially the same and interpretable. They are both manifestations of energy release and absorption on the locked fault during the meta-instability stage, except that the feature shown in Figure 4 (a) is indirectly caused.

## 5.3. Relationship

The contour maps of the apparent wave velocity ratio and apparent Poisson's ratio, along with the divergence map of their gradients, collectively reflect the stress field state of the fault near the rupture earthquake. Specifically, the trend of the fitting line of the minimum value points of the apparent velocity ratio (142.8°) and the fitting line of the maximum value points of its gradient divergence (141.8°), as well as the trend of the fitting line of the minimum value points of the apparent Poisson's ratio (143.9°) and the trend of the fitting line of the maximum value points of its gradient divergence (144.0°), are consistent with the trend (138°) of the node 1 of the earthquake in the CMT product catalog of the focal mechanism of China Earthquake Networks Center (https://data.earthquake.cn). Additionally, the trend of the fitting line for the maximum points of the apparent wave velocity ratio (40.1°) and the minimum points of its gradient divergence (39.1°), as well as the trend of the fitting line for the maximum points of the apparent Poisson's ratio (34.2°) and the minimum points of its gradient divergence (49.9°), are consistent with the trend of node 2 (45°) (although there are some differences, they are considered to be within an understandable range). This consistency fully demonstrates the precursor property, that is, the state of the stress field before the instability and rupture of the fault. The energy absorption and release reflected by the divergence of gradients are consistent and unified with the changes in apparent wave velocity ratio and apparent Poisson's ratio, which are influenced by the temperature changes of the medium during this process. Given that earthquakes are a phenomenon of energy released by fault rupture, the brewing process reflects the overall stress field movement in the region. Therefore, anomalies in local data do not necessarily have complete precursor significance, similarly, anomalies in several different types of data may also be unable to make effective judgments on precursors due to the lack of organic connections. Instead, the movement of fields related to the stress field should be regarded as a whole to evaluate the state of the stress field and effectively judge its precursors. The geographical distribution of locked faults can be estimated based on the extreme points of energy release. The specific epicenter location may be affected by the constantly changing geophysical environment on the locked fault. From an energy perspective, the probability of unstable rupture points being located in areas with greater energy release may be higher.

It can be inferred that the possible process of this earthquake is as follows: In the interaction between the blocks, the stress field is influenced by the fault zone in the region, and the energy on the locked fault continues to accumulate. When the strain energy on the locked fault accumulates to the rock-bearing limit, it activates the locked fault and enters the irreversible meta-instability stage of rupture. The locked fault changes from mainly absorbing energy to mainly releasing energy. With the release of energy on the locked fault and the absorption of energy on its conjugate surface, the energy conversion becomes more intense. As it further develops and synergizes to a certain extent, the extreme points of energy release shows signs of activity on the locked fault. Simultaneously, the release and absorption of energy in rock media cause the temperature at corresponding locations to decrease and increase, resulting in minimum and maximum values of wave velocity ratio and Poisson's ratio (the apparent wave velocity ratio and apparent Poisson's ratio on the vertical ground also exhibit minimum and maximum values). With the intensification of stress field activity, the synergy becomes increasingly strong, and the fault eventually becomes unstable, ruptures and earthquakes occur.

Based on the above analysis, it is believed that the meta-instability precursor of the Yangbi $M_S$6.4 earthquakes can be understood as the characteristic of the stress field state before the rock medium becomes unstable and fractures, as well as the physical response to the surrounding medium during the energy transfer process. This finding is not only consistent with the definition of precursors described in the literature (Chen YT, 2009) but also verifies the relationship between seismic wave velocity, faults, and stress factors described in experiments. It also conforms to the characteristic of mutual confirmation of different physical fields in meta-instable states (Ma J et al., 2016). Additionally, during the processing, although the apparent wave velocity ratio and apparent Poisson's ratio have the same error impact, they have a nonlinear relationship, and the consistency of their contour lines and gradient divergence distribution characteristics also confirm the objectivity of their mathematical relationship. In field observations, it is believed that this method has certain reference significance for identifying meta-instable stages and discovering precursors.

### 5.4. Question and future

The effectiveness of this method largely depends on the richness of the data and is constrained by the geometric relationship between the observation plane and the stress field. Specifically, the state of the stress field can only be effectively captured when the observation plane is orthogonal to the stress field. In non-orthogonal cases, the vertical projection of the stress field onto the ground may become indistinguishable due to overlap. Additionally, the presence of minerals such as oil and gas in rock media may cause discontinuities in the stress field, further complicating the interpretation of its projection on the ground and posing significant challenges to the study of the meta-instable state. In the case of Yangbi $M_S$6.4 earthquakes, the rupture fault of the shallow strike-slip type (Yang JY et al., 2021), which is orthogonal to the stress field, this orthogonality allows the vertical projection on the ground to effectively reflect the state of the stress field. Although the above situation restricts further research on the stress field, the vertical projection of the stress field on the ground still provides valuable insights into the stress field state within the earthquake depth range. Therefore, utilizing data with precise depth can help capture detailed dynamic changes in the stress field from different depth vertical projections in three dimensions, offering a valuable reference for earthquake prediction and disaster prevention.

## 6 Conclusion

(1) The precursor information of meta-instability is a comprehensive reflections of the stress field state and its associated physical phenomena in the rock medium before the fault rupture. Due to the complexity of rock media, there may be some differences in the physical forms reflected by different observation methods.
(2) In field observations, data interpolation is an effective method for discovering precursor information, particularly in areas with sparse data.
(3) Quantities related to seismic wave velocity can effectively reflect the state of the stress field, with their gradients and divergence patterns showing consistency.
(4) The identification of the meta-instable stage relies on a substantial amount of high-quality data to ensure the accuracy and reliability of the analysis results.

(5) When the two-dimensional plane of the Earth's surface is orthogonal to the stress field, meta-instability precursor information is easier to identify. However, this situation may be influenced by the continuity of the underground medium. In non-orthogonal cases, projection overlap complicates the interpretation of surface observation, increasing the difficulty of identifying the stress field state.

## Acknowledgements:

We would like to express our gratitude for using GMT (Wessel and Luis, 2017) and the MATLAB-based M-map (https://www.eoas.ubc.ca) toolbox in our calculations and graphing!

## Conflict of interest:

The authors affirm that they have no financial and personal relationships with any individuals or organization that could have potentially influenced the work presented in this paper.